\documentstyle[12pt]{article}
\normalsize

\def\a{\alpha}

\def\o{\omega}
\def\p{\pi}

\def\r{\rho}

\def\D{\Delta}

\def\G{\Gamma}

\def\L{\Lambda}

\def\rt{\rightarrow}

\def\bar#1{\overline{#1}}

\def\Hat#1{\rlap{\kern.10em$\widehat{\phantom G}$}#1}
\def\HAt#1{\rlap{\kern.05em$\widehat{\phantom G}$}#1}

\def\cap#1{\rlap{\kern.1em$\widehat{\phantom{G\vrule height.8em}}$}#1{}}
\def\Cap#1{\rlap{\kern.05em$\widehat{\phantom{G\vrule height.8em}}$}#1{}}

\let\oldtheequation=\theequation
\def\doteqs#1{\setcounter{equation}{0}
            \def\theequation{{#1}.\oldtheequation}}
\newcounter{sxn}
\def\sx#1{\addtocounter{sxn}{1} \bigskip\medskip \goodbreak
\noindent{\large\bf\centerline{\thesxn.~~#1}} \nobreak \medskip}
\def\sxn#1{\sx{#1} \doteqs{\thesxn}}

\newcounter{axn}

\def\br{\begin{thebibliography}{abc}}
\def\er{\end{thebibliography}}

\def\be{\begin{equation}}
\def\ee{\end{equation}}
\date{}

\begin{document}
\bibliographystyle{unsrt}
\footskip 1.0cm
\thispagestyle{empty}
\setcounter{page}{0}
\begin{flushright}
UR-1326\\
ER-40685-776\\
SU-4240-557\\
February 1994\\
\end{flushright}
\vspace{10mm}
\centerline{\large \bf SEMICLASSICAL DECAY OF EXCITED STRING STATES}
\vspace{5mm}
\centerline{\large \bf ON LEADING REGGE TRAJECTORIES}
\vspace{5mm}
\centerline{\large  K. S. Gupta$^*$, C. Rosenzweig$^{\ddag}$}
\vspace*{5mm}
\centerline {$^*$\it {Department of Physics and Astronomy, University of
Rochester},}
\centerline {\it Rochester, NY 14627, USA}
\vspace{5mm}
\centerline {$^{\ddag}$\it {Department of Physics, Syracuse University},}
\centerline {\it Syracuse, NY 13244-1130, USA}
\vspace*{25mm}
\baselineskip=24pt
\setcounter{page}{1}

\centerline { \bf Abstract}

	We study the decay of hadrons based on a semiclassical string model.
By including quark mass effects  we find that the
width to mass ratio $\G/m$ is an
increasing function of $m$, which increases most rapidly for massive quarks.
This  is consistent with the available data. The decay probability of hadrons
on the leading Regge trajectories is computed taking the effect of the string
rotation into account. The resulting decay probability is no longer uniform
along the length of the string but varies in a manner that is in qualitative
agreement with the available data. We argue in favour of possible
experiments that would test our predictions more accurately and help open a
window to the nonperturbative aspects of QCD.

\vfill
\newpage
\sxn{Introduction}

Much of the physics of confinement can be understood on the basis of a
simple string picture of hadrons.  In particular, a linear potential between a
quark and an antiquark and a linearly rising Regge trajectory are simple,
immediate consequences of this picture.  The linearity of Regge trajectories
is,
in fact, the most striking evidence that we have for the string model.  The
spin $\bf{6}$  meson of mass 2450 which lies on the (still very) linear $\rho -
A_2$ trajectory is dramatic evidence for the existence of a string out to
distances of almost three fermis! Each newly discovered high spin meson on this
trajectory extends our knowledge of strings by about 1/4 fermi.  The
reproduction of
linear potentials was one of the earliest and most encouraging successes of
lattice gauge theories.  The string model has successfully been applied to the
hadronization of quarks produced in $e^+$ $ e^-$ collisions.  The string
model is useful, fecund and well established.

Study of the hadronic string model gave rise to superstring theory where truly
 one
dimensional strings are (The) fundamental objects.  It is clear that QCD
strings are not one dimensional but have finite extent like flux tubes
and thus cannot be treated by fundamental theory.  Nevertheless they are still
predominantly stringy and therefore we may learn  about them from studying
superstring theory.  Recently, there has been much interest in studying the
relationship  between fundamental string models and the three dimensional flux
tubes connecting physical quarks.$^1$

The success of the string picture and its physical evocativity behooves us to
explore how far we can extend the simple, semi-classical picture of hadrons.
Since the existence of linearly rising Regge trajectories is at once the main
test and main manifestation of the string picture it pays to see what further
physics can be extracted from this simple physical system.  The essence of the
string model for hadrons is the relationship between the energy (mass) and the
angular momentum of a spinning string (or quasi one dimensional object such as
a flux tube). The energy and the angular momentum are given by
\begin{equation}
E=\frac{\pi k}{\omega},  ~~~~~~~~J=\frac{\pi}{2}~ \frac{k}{\omega^{2}},
\end{equation}
where
$\omega$ is the angular frequency of the string and $k$ is the string tension.
Finite mass quarks can be
placed at the ends with interesting implications.$^{2-4}$

The string model becomes semi-classical when we account for its breaking.  In
the presence of a large chromo electric field a $q-\bar{q}$  can tunnel out of
the vacuum and eat up some of the string.$^5$  This process has been studied
since
the earliest days of QED and, at least for electrodynamics, is well
understood.$^6$  The string will break with equal probability per unit four
volume.
The string breaking probability for QED is
\begin{equation}
p=\frac {\alpha E^{2}}{2\pi^{2}} \sum_{n} \frac{1}{n^{2}} \exp ~
\left (\frac {-n\pi m^{2}}{|eE|} \right ),
\end{equation}
\noindent where $m$ is the quark mass, $k=eE$ is the string tension expressed
in terms of the electric field $E$.  Heavier quarks are very much less likely
to pop out of the vacuum.
This string breaking picture has been successfully employed in fragmentation
models which provide qualitative support for the ansatze.$^7$

 For a string state the total width is proportional to the total decay
probability.  In the semiclassical string breaking model this probability is
directly proportional to the string length.  On the leading Regge trajectory,
since the mass is proportional to the length, the picture predicts the ratio of
width to mass, $\G/m$, should be constant. For mesons in the particle data
book the
width to mass ratio ranges from .05 to .2, providing at best only weak
qualitative agreement.  As we shall see some of the trends inherent in this
data can be explained by the more detailed picture we develop below.

In Section II we consider the corrections to the simplest string model by
allowing for massive quarks at the ends of the strings. For massive quarks
a state of {\em given} energy will have less of its energy in the string than
if
the quarks were massless. Thus the length of string, and hence its decay
probability will be smaller than it would be for the massless case.
Asymptotically as the hadron mass increases (and the quark masses become
negligible with respect to the string energy) $\G/m$ will increase to its
constant universal value. Nevertheless for
 light mass quarks ($m<$200-300 MeV) the trajectories will remain
nearly linear.  The massive quark takes up almost exactly the energy {\em and}
angular momentum of the string it replaces maintaining an almost linear
relationship between J and E.

The leading Regge trajectory comes about because the string is spinning.  This
will introduce centrifugal effects which will change the simple minded
prediction.  In Section III we model the centrifugal effects and find that
strings prefer to decay symmetrically to states of angular momentum $\sim J/2$
rather than decaying by cascading down to a state with nearby $J$.  These
centrifugal effects also cause the decay probability for particles on the Regge
trajectory to be different than that used in the string fragmentation models.

It strikes as unfortunate that because of the rush to higher energy physics the
field of ``low energy'' Regge spectroscopy has languished with almost no
progress in the last ten years.  Yet it appears that this is the most
productive
place to experimentally explore the details of the confinement mechanism of
QCD.  Surely this is a worthy topic for extended effort even as we test
the (basically perturbative) predictions of the Standard Model to ever higher
precision.  There are many important conceptual questions about the nature of
confinement that experimental data can elucidate.  For instance, in fundamental
string theory there is evidence that the width of the particles on the leading
trajectory will level off with increasing mass (and $J$) whereas the naive
physical flux tube model predicts a linear increase.$^8$  Experimental studies
providing more accurate widths and much higher spin states would distinguish
between these predictions and could thus provide clues as to the relationship
between fundamental strings and semi-classical, thick strings.  In general any
detailed information about the interactions and breakings of
clearly delineated, extended strings must be useful.  We urge our experimental
colleagues to revive this once active field of particle physics research.

\sxn{Classical Corrections:  Finite Mass Quarks}

When we place massive quarks at the ends of the string the simple relationship
Eq. (1.1) between mass and angular momentum is modified and becomes a
parametric
relationship :
%

\begin{equation}
M = \frac {m_{1}}  {(1-v_{1}^{2})^{\frac{1}{2}}}~ + ~\frac {m_{2}}
{(1-v_{2}^{2})^{\frac{1}{2}}}\nonumber
+ \frac {k} {\omega} ~\int_{-v_{2}}^{v_{1}} ~
\frac {dv}  {(1-v^{2})^{\frac{1}{2}}},
\end{equation}
\begin{equation}
J = \frac {m_{1}v_{1}^{2}}  {\omega (1-v_{1}^{2})^{\frac{1}{2}}} ~ + ~
\frac {m_{2}v_{2}^{2}}  {\omega(1-v_{2}^{2})^{\frac{1}{2}}}\nonumber
+ \frac {k}{\omega^{2}} ~\int_{-v_{2}}^{v_{1}} dv
\frac {v^{2}}  {(1-v^{2})^{\frac{1}{2}}},
\end{equation}
\begin{equation}
v_i=\left( 1+\left (\frac{m_i\o}{2k}\right )^{2}\right )^{\frac{1}{2}}
-\frac {m_i\o}{2k},
\end{equation}
where $v_i$ are the velocities of the quarks of masses $m_i$.

Intuition suggests that the string will shrink and may slow its rotational
velocity compared to the case of massless quarks at the ends.  To verify this
we consider the case when the velocity of the massive quark is relativistic.
This occurs when $ \frac {m\o}{2k}$ is small.  By using Eq. 2.1-2.3 we find the
leading corrections of order
$ \left (\frac{m\o}{2k}\right )$ cancel
out implying that the relationship between $J,M$, and are very similar to
the massless case :

$$
E=\frac{k}{\o} \left ( \p + \frac{4\sqrt{2}}{3} \left ( \frac{m\o}{2k}\right)
^{\frac{3}{2}}\right ),
$$
\begin{equation}
J=\frac{k}{\o^{2}} \left( \frac {\p}{2} - \frac
{4\sqrt{2}}{3} \left ( \frac {m\o}{2k}\right )^{\frac {3}{2}}\right ).
\end{equation}

Neglecting terms of order $\left (\frac{m\o}{2k}\right )
^{\frac{3}{2}}$ these are formally identical to the massless case.  To order
 $\frac {m\o} {k}$,  for fixed $J$ (or $E$), $\o$ and $E$ (or $J$) are
unchanged.  The velocity of the endpoint, and hence the string length $v=\o L$
decreases by $\frac {m\o} {2k}$.   The relationship Eq. 1.1  is now

\begin{equation}
J=\frac{E^{2}}{2\pi k} - \frac {4}{3} \left (\frac{2}{\p}\right )
^{\frac{1}{4}}
\left (\frac{m}{\sqrt{k}}\right )^{\frac {3}{2}} J^{\frac {1}{4}}.
\end{equation}

The correction to linearity is both relatively small and slowly varying with
$J$.  It can
easily be masked when including the Regge intercept (or quantum defect)
$J=1/(2\p
k)E^2+\a_0$ in a fit to the Regge trajectory. $^{3-4}$

When finite mass quarks are at the ends, the length of the string, to which the
semiclassical width is proportional, increases faster than the energy with
increasing $E$.  The string grows ``faster'' than in the massless case.  For
small mass, and hence $\frac {m \o}{2k}$ small, we find (keeping only terms of
order $\frac {m \o}{2k}$) that
\begin{equation}
\frac {L}{E} =\frac{2}{k \p} -
\left [ \frac {(m_{1}+m_{2})}
{2kE} \right ].
\end{equation}

This changes the naive prediction that the ratio of the width to particle mass
should be a universal constant for particles on leading Regge Trajectories.  To
compare to experiment we must chose reasonable values for the masses.

In a previous fit to the Regge trajectories of the strange and the nonstrange
mesons the values $m(u,d)=175$ and $m(s)=345$ were selected [4]. We use these
values to compute the string length for the corresponding mesons. In Fig. 1 we
display the ratio of the length to the energy of the $K^*$ trajectory as
calculated from Eq. 2.1 - 2.3 with
$\a' =.92$.  We have multiplied the ratio $L/E$ by
1/29 so as to normalize this ratio to the observed ratio of the $K^*(3)$ width
to its mass.  We remind the reader that this curve is not a fit but a
prediction
(up to normalization) from the parameters of Ref. 4. Fig. 2 is the
corresponding plot for the $\r-A_2$ trajectory.  As the experimental points we
plot the average value of $\G/m$ for the $I=1$ and the $I=0$ particles. The
string
model does not distinguish between different isospin states and so this is the
natural object to compare to.  Experimentally the small pion mass and $G$
parity play a large and possibly distorting role in the decay patterns for the
low mass states and the averaging is a reasonable way to eliminate the wide
variations.  Had we not done so the data are patternless and reflect the mass
and quantum number constraints.  The dramatic success of the $K^*$ trajectory
is vitiated by the poor correspondence in the $\r-A_2$ case.    Nevertheless,
even for this case, there is evidence
for a gradual increase.  (We have used the same normalization as for the $K^*$
trajectory as would be required by the string model.)  The
relatively sharp increase of the ``predicted'' curve is
driven by the mass of the light quark.  The data can be reproduced by a smaller
mass of about 50-100 MeV for $m(u,d)$.  The point we want to emphasize here is
that the
meson data show the width increases faster that the mass, markedly so for
the strange mesons and much less so for the $\r-A2$ trajectory.  The string
model with finite quark mass, qualitatively, predicts just this trend. We can
also apply the model to high spin baryons on the leading baryon Regge
trajectories.  Here too linearity of the Regge trajectory implies a string
model.  One of the three quarks is excited producing a string like structure
connecting it with the two remaining quarks (diquark$^9$).  By appropriate
choice of diquark mass, $m$ = .975, the string with massive quarks at the end
readily
accommodates the trend in $\G/m$.  See Fig. 3.

\sxn {Quantum  Corrections: Centrifugal Barriers to Tunnelling}

A simple model for string breaking involves quark pair creation by the strong
chromoelectric field inside the string.  Although a field theory treatment is
required for a correct calculation the essential physics is contained in a
semiclassical analysis of tunneling, from the vacuum, of a quark antiquark
pair.
Since we are looking at semi-quantitative results this semiclassical picture
will be sufficient.  Creation of massive $q-\bar {q}$ from the vacuum in a
relativistic effect and so we employ the Klein-Gordon equation in the WKB
amplitude for tunneling.  In this treatment we are ignoring the spin of the
quarks so that the Klein Gordon equation should be sufficient.  A quark of
mass $m$ pops out of the vacuum materializing after vaporizing some of the
string and  leaving behind an antiquark hole.  On its way out it experiences a
$kr$ linear potential.  The WKB expression for the tunneling event is
\begin{equation}
P \sim \exp \left ( -2 \int_{0}^{r_{c}} \sqrt{(E-V(r))^{2}-m^{2}}~dr \right )
= \exp \left (-\frac {\pi m^{2}}{k} \right ).
\end{equation}
\noindent $r_c$ is the classical turning point, and $E=-m$.  Remarkably this
is identical to the exact field theory result of
Euler, Heisenberg, and Schwinger.$^5$

We are interested in how this simple, one-dimensional, result changes when,
as is
appropriate for particles on Regge trajectories, the string is rapidly
rotating.  The quark will now have to overcome a centrifugal barrier to tunnel
out.  This can be accounted for by using $WKB$ for the  radial Klein Gordon
equation.

\begin{equation}
P \sim \exp \left (-2 \int_{0}^{r_{c}} \sqrt {(E-V(r))^{2}-m^{2}
-\frac{l(l+1)}{r^{2}}} \right ).
\end{equation}

\noindent The $l$ appearing here is the angular momentum of the string picked
up by the quark as it tunnels out.  Strictly speaking we should replace $l$
$(l+1)$ by $(l+\frac {1}{2})^2$ incorporating the Langer correction.$^{10}$
This
correction emphasis the region near $r=0$ (as we show below $l \rt 0$ as $r\rt
0)$ which is where our ignorance of the physics is greatest and will force us
to
introduce an extra parameter as a cutoff.  Since we do not expect our results
to be quantitative we chose to continue with the  uncorrected centrifugal term
rather than add another, uncontrolled parameter.  This should still be a
reliable qualitative guide to the centrifugal effects.  We thus need to know
the angular momentum carried by the vaporized string.

Consider a point a distance $R$ away from the pivot point B (center of mass)
of a  string
rotating with angular frequency $\o$ (see Eq. 2.1-2.3).  Due to the string's
rotation this point moves with a velocity $v=\o R$ perpendicular to the string.
Consider a quark tunneling from $R$ to a point $R+r$ away from the pivot point
(see Fig. 4., where B is the center of mass of the whole string AE before
breaking and CD is the portion that vaporizes.)
The quark acquires the angular momentum of the vaporized string.  This angular
momentum, relative to the point $R$ is
\begin{equation}
l(r) = \int_{R}^{R+r} \frac {kd\r v (\r)\r} {\sqrt{1-v(\r)^{2}}},
\end{equation}

\noindent where $v(\r)$ is the relative velocity of the string element with
respect to $R$ :
\begin{equation}
v(\r)= \frac {\o\r}{1-\o^{2}(\r+R)R} ~.
\end{equation}

\noindent In many situations of physical interest the value of $r$ is small
and a
small $r$ expansion is adequate.  This yields
\begin{equation}
l(r)=\frac {k\o} {(1-\o^{2}R^{2})} \left [
\frac {r^{3}} {3} +
\frac {R\o^{2}} {(1-\o^{2}R^{2})} \frac {r^{4}} {4} +~~...\right ].
\end{equation}
\noindent  When this approximation is valid we can neglect the quadratic term
$l^2$ in the tunneling amplitude and find for the string breaking probability
for a spinning string
\begin{equation}
P\sim \exp - \frac
{m^{2}\pi}{k}
\left \{ \frac {\left (1+
\frac
{{w}}
{6m(1-\o^{2}R^{2})}
\right )^{2}}
{\left (1 -
\frac
{w^{3}R}
{4k(1-\o^{2}R^{2})^{2}}
\right )^{3/2}}  \right \}.
\end{equation}
\noindent Compared to the one dimensional case there is an additional factor
in the curly
brackets in the exponent of the decay probability.  This extra suppression is
position dependent, as we might suspect for a spinning string.  Tunneling is
more difficult near the end of the string than at the fulcrum.  For instances
for a high angular momentum string with $\o$=.05 and $m$=.1 ($k$=1/5.8) the
probability is 30\% higher for the string to break at the center ($R$=0) than
near the end ($R$=18).  This has several interesting phenomenological
consequences in the semiclassical string model.  First we see that the string
is more likely to break near its middle in a symmetric way into two high
angular momentum states rather than splitting near the end of the string and
cascading down by the emission of a ``pion'' to a high spin precursor.  The
semiclassical string model quantifies this simple physics conclusion by
providing a specific recipe for the centrifugal suppression.  This type of
suppression is consistent with the sparse data that exists.

The semiclassical string model has seen its widest usage as an input for the
so-called Lund Fragmentations model.  In that situation there is no need to
take into account the centrifugal effects because we believe the process is
one dimensional to a high degree of accuracy.  An input to this process is the
decay probability per unit length of the string.  This value is usually assumed
to be the same as that deduced from the widths of particles on the leading
Regge trajectories.  However because of the centrifugal effects calculated
above these two quantities are not the same.   Neither the string
fragmentation model nor
the Regge string decay model is so precise that this is significant but it is
good to keep in mind when comparing ever sophisticated versions of the two
semiclassical string models.

As the quarks tunnel out from the vacuum they eat up some of the string.  This
restricts the amount of string available for decay.  For instance (ignoring the
centrifugal effects) the quarks  require a minimum length of $2m/k$ in order to
tunnel out.  Similarly the string cannot break within a distance of $2m/k$ of
the ends.  This further modifies the simple relation between width and length
of the string as only a part of the string is available for decay. Including
the centrifugal effects will require that even more string must vaporize in
order for the quarks to materialize.  The length of the vaporized string
depends on its position on the string via the $R$ dependence appearing in Eq.
(3.3).  Since more string is required for larger $R$ this effect will further
suppress non symmetrical decays with respect to the symmetrical decays.  The
$R$ dependence in Eq. (3.3-3.6) is not dramatic however and so there will be
only mild position dependence.

Phenomenologically the simplest approach to these effects is to set the $\G/m$
ratio equal to $L-\D$ ignore the $R$ dependence and consider $\D$ as a
parameter.  This provides the model with sufficient robustness to accommodate
the data incorporated in Fig. 1-3.  We return to this data and relax the
requirement that the masses are given by the best fit to the Regge trajectory.
We still restrict the quark masses to the range found acceptable [4] for
reproducing Regge trajectories.  Using 75 MeV for the light quarks, 400 MeV for
the strange quark,  1000 MeV for the  $\L$ diquarks and a
value of .2 GeV$^{-1}$ for $\D$ we can easily represent the decay data.  This
is
displayed in Figures 5-8.  The quark  masses are
reasonable but somewhat smaller than usual constituent quark masses while the
$\L$ diquark mass is on the high side.  The value of $\D$ is smaller than we
might have expected from the above physical discussion based on semiclassical
tunneling. The nulceon data also show the expected trend in $\G/m$ but the
uncertainties in $\G$ for high $J$ states is so large that a graph of $\G/m$
vs. $L/E$ is not illuminating.  It is quite clear that the simple, but
physical, semiclassical
string model provides a good description of the total decay widths of particles
on the leading meson and baryon Regge trajectories.

The relevance of the semiclassical picture is satisfyingly supported by the
numerical comparison of the phenomenologically extracted string decay
probability with the most naively calculated value.  Eq. (1.2) is the
semiclassical decay probability per unit four volume.  To convert this to
the probability  of string decay we must multiply by the cross sectional area
of the string,  a factor of two for the quark spin and a factor of at least two
for the two (or three?) flavors of light quarks that can independently pop out
of the vacuum.  Taking a flux tube radius of $\sim$1/300 MeV and using
$k=1/5.8$, $m=.075$ the semiclassical result 1/65 is comparable in magnitude to
the 1/27.5 used in Figures 5-8.  Given the myriad of uncertainties in the model
and the values of the parameters (especially the flux tube radius) this
confluence is most encouraging.

We have not displayed our model for the charm
meson trajectory because of its embarrassing inadequacy.  The $D^{**}$ meson,
while sitting appropriately on a charm meson trajectory [4], is exceedingly
narrow, $\G\sim$ 25 MeV.  This is somewhat troublesome for any meson model,
certainly our own.  In the simplest picture we expect heavy quark mesons to be
narrower than their light quark counterparts because the string, for the same
$J$, is shorter for the heavier meson (by a factor of $\sqrt{2}$ in the
infinite
mass heavy quark limit).  The tunneling effects discussed above and embodied in
the parameter $\D$ emphasize this decrease.  For instance, there is no
``symmetric'' decay possible for a heavy quark meson.  Still the width does
seem inordinately small.  Nevertheless the width of the first excited state on
a heavy meson trajectory is very sensitive to these effects and we can easily
imagine that a more accurate treatment of them will produce the observed width.
As higher excitations are discovered we expect the widths to approach the
typical widths of the other mesons.

\sxn {Conclusions}

The string model provides a beautiful, physical picture of quark confinement.
It strongly motivates the linear part of the quark anti quark potential which
has met with great success in  heavy quark systems.  Strings predict and
account for the remarkable linearity of both meson and baryon Regge
trajectories.  In this paper we went beyond the spectroscopic aspects of the
string model and explored the implications of string breaking for hadron
decays.  The  mechanism for breaking is envisioned as the tunneling from the
vacuum of $q-\bar {q}$ pair.  This leads to the prediction that the ratio of
the hadron width to its mass is a universal constant, the string breaking
probability.  If we take isospin averages, (in order to average over phase
space effects due to the pion mass) the data are in rough agreement with the
expectation.  A definite deviation is, however, evident.  The width to mass
ratio is an increasing function of the particle mass.  This trend is, in fact,
exactly what the most immediate refinements to the naive model predict.
Quarks  have mass and when placed at the ends of strings they shorten the
string length for constant mass hadron.  Thus $\G /m$ while still proportional
to $L/m$ is now a rising function of $m$ which goes to a universal constant
asymptotically.  The fact that tunneling requires a finite length of string
reinforces this trend.  Indeed the data are consistent with the combination of
these two effects and thus supportive of semiclassical string picture of decay.

Our major results are

\noindent 1) To first order in quark mass, massive quarks shorten
the length  of the string but do not alter the linearity of the Regge
trajectory.
This explains the dramatic success of the massless quark string model for Regge
trajectories.

\noindent 2) When we isospin average the meson widths, the width to mass data
occupy a narrow range from .055 o .15 and show a definite trend of increasing
with the mass of the hadron.

\noindent 3) Straightforward incorporation of finite quark mass effects and a
semiclassical treatment of decay reproduce this trend in the data.

\noindent 4) The ``experimental''  value of the string decay constant is very
broadly consistent with the theoretical tunneling probability.

\noindent 5)  The centrifugal barrier to tunneling gives rise to a preference
for the string to decay symmetrically.  This is consistent with an experimental
preference for such decays and with decays of highly excited fundamental string
in 26 dimension but, possibly, not consistent with fundamental string decays in
less than the critical dimension.

\noindent 6) Because of the positional dependence of the centrifugal barrier
strictly speaking the string decay consistent is different for states on the
Regge trajectory from that for use in fragmentation functions.

Our results demonstrate the continued vigor of the string model and its
robustness in confronting and illuminating the dynamical process of hadron
decay.  We hope it provides encouragement for future studies, both experimental
and theoretical, of hadronic strings.  The problem of understanding quark
confinement as a consequence of the QCD Lagrangian is a formidable one that has
so far resisted twenty years of vigorous attack.  We are however, possessed of
a
physically compelling model that has an excellent chance of emerging from the
final solution to the confinement problem.  Let us fully exploit it.

\vfill
\newpage
\centerline {\bf Acknowledgements}
\vglue 0.5cm

	We thank P. Teotonio for discussions. K.S.G
was supported in part by the US Department of Energy, Grant
No. DE-FG02-91ER40685. C. R. was supported by the Department of Energy under
contract number DE-FG02-85ER40231.

\vglue 0.6cm

\centerline {\bf References}
\vglue 0.5cm
\begin{enumerate}
\item J. Polchinski and A. Strominger, {\it Phys. Rev. Lett.} {\bf 67}, 1681
 (1991). J. Polchinski, {\it Phys. Rev. Lett.} {\bf 68}, 1267 (1992). B.S.
Balakrishna, {\it Phys. Rev.} {\bf D48}, R5471 (1993)
\item A. Chodos and C. Thorne, {\it Nucl. Phys.} {\bf B72}, 509 (1974).
\item K. Johnson and C. Nohl, {\it Phys. Rev.} {\bf D19}, 291 (1979).
\item F. Lizzi and C. Rosenzweig, {\it Phys. Rev} {\bf D33}, 1685 (1985).
\item A. Casher, H. Neuberger and S. Nussinov, {\it Phys. Rev.} {\bf D20} 179
(1979).
\item C. Itzychson and J.B. Zuber, {\it Quantum Field Theory}, McGraw Hill, NY,
1980, Chapter 4.
\item B. Andersson, G. Gustafson and C. Peterson, {\it Z. Phys. C.} {\bf 1},
105
(1979). M.G. Bowler, {\it Z. Phys. C} {\bf 11} 169 (1981).
\item D. Mitchell, N. Turok, R. Wilkinson, and P. Jetzer, {\it Nucl. Phys.}
{\bf B315} 1 (1989). D. Mitchell, B. Sundberg, and N. Turok, {\it Nucl. Phys.}
{\bf 621}, (1996).
\item M. Anselmo et al. {\it RMP} {\bf 65}, 1199 (1993).
\item R.E. Langer, {\it Phys. Rev.} {\bf 51}, 669 (1937).

\end{enumerate}

\vfill
\newpage
\centerline {\bf Figure Captions}
\vglue 0.5cm

\begin{itemize}
\item [\bf {Fig. 1}]  The isospin averaged $\G/m$ for particles on the $\r
-A_2$ trajectory plotted on the same plot as $L/E$ for the corresponding string
model states.  The experimental $\G/m$ has been multiplied by 29.

\item [\bf {Fig. 2}]  Same as Fig. 1 for $K^*$ trajectory.

\item [\bf {Fig. 3}]  Same as Fig. 1 for $\L$ trajectory.

\item [\bf {Fig. 4}] String after tunneling of $q \bar {q}$ and consequent
vaporization of
a string segment.

\item [\bf {Fig. 5}] String model, as explained in text, for $\r-A_2$
trajectory. $\G/m$ is multiplied by 27.5.

\item [\bf {Fig. 6}] Same as Fig. 5 but for $K^*$ trajectory.

\item [\bf {Fig. 7}] Same as Fig. 5 but for $\phi$ trajectory.

\item [\bf {Fig. 8}] Same as Fig. 5 but for $\L$ trajectory.

\end{itemize}
\end{document}